\documentclass[mathleft,finallayout
]{an}
\usepackage{graphicx}
\usepackage{times}
\begin{document}

\Pagespan{871}{874}
\Yearpublication{2007}%
\Yearsubmission{2007}%
\Month{10}%
\Volume{328}%
\Issue{8}%
\DOI{10.1002/asna.200710809}%

\title{Molecular cloud abundances and anomalous diffusion}

\author{G. Marschalk\'o$^{1,}$\thanks{Corresponding author:
  \email{G.Marschalko@astro.elte.hu}\newline}
\and  E. Forg\'acs-Dajka$^{1}$
\and  K. Petrovay$^{1,2}$
}
\titlerunning{Molecular abundances and anomalous diffusion}
\authorrunning{Marschalk\'o, Forg\'acs-Dajka and Petrovay}
\institute{
$^1$E\"otv\"os University, Department of Astronomy, Budapest, Pf.~32, 
H-1518 Hungary\\
$^2$ASIAA/National Tsing Hua University - TIARA, Hsinchu, Taiwan
}

\received{}
\accepted{}
\publonline{}


\abstract{
The chemistry of molecular clouds has been studied for decades, with
an increasingly general and sophisticated treatment of the reactions involved.
Yet  the treatment of turbulent diffusion has remained extremely sketchy,
assuming simple Fickian diffusion with a scalar diffusivity $D$. However,
turbulent flows similar to those in the interstellar medium are known to give
rise to anomalous diffusion phenomena, more specifically superdiffusion
(increase of the diffusivity with the spatial scales involved). This paper
considers to what extent and in what sense superdiffusion modifies molecular
abundances in interstellar clouds. For this first exploration of the subject we
employ a very rough treatment of the chemistry and the effect of non-unifom
cloud density on the diffusion equation is also treated in a simplified way. The
results nevertheless clearly demonstrate that the effect of superdiffusion is
quite significant, abundance values at a given radius being modified by order of
unity factors. The sense and character of this influence is highly nontrivial.}

\maketitle

\section{Introduction}

The chemistry of molecular clouds is a complex system affected by different
factors.  The first models laid down the foundations of the chemical reaction
network: following the initial steady state gas-phase chemistry model of Herbst
\& Klemperer (1973) several (pseudo-)time-dependent models, with fixed profiles
of physical parameters were developed, considering more and more chemical
reactions, while neglecting photodissociation (Leung et al.~1984; Herbst \&
Leung 1989; Millar \& Herbst 1990). The agreement of the mo\-dels with observed
fractional abundances improved when the effect of turbulent diffusion was taken
into account, resulting in smoother density distribution profiles for the more
important species (Xie et al.~1995, hereafter XAL95). Subsequent models further
refined the chemistry, considering details of the ion-neutral reaction scheme,
and taking into account the gas-grains interaction, $\mathrm{H_2}$ and CO
self-shield\-ing, grain accretion effects and adsorption onto grains (Will\-acy
et al.~2002; Yate \& Millar 2003).

In contrast to the enormous efforts made in order to improve the representation
of chemical processes, the treatment of turbulent diffusion has remained
extremely sketchy, assuming simple Fickian diffusion with a scalar diffusivity
$D$. Under the conditions prevailing in interstellar space this is surely wrong.
Interstellar turbulence is mainly driven by supernova shocks on spatial scales
far exceeding that of the molecular clouds. From those scales, kinetic energy
cascades down to the scale of clouds, cloud clumps and even cloud cores
(Boldyrev 2002; Scalo \& Elmegreen 2004; Ryan Joung et al.~2006). Indeed, in
current thinking clouds are hardly more than positive density fluctuations in
this large-scale compressible turbulent flow. In such a turbulent flow, the
increase of the r.m.s. separation $\Delta$ of two tracer particles will only be
due to eddies smaller than $\Delta$, leading to a {\it scale-dependent eddy
diffusivity.} As a result, the increase of $\Delta$ with time will deviate from
the Fickian square-root law: such {\it non-Fickian} or {\it anomalous} diffusive
pro\-cess\-es commonly occur in various fields of physics (Avellaneda \& Majda
1992), though astrophysical applications have so far been limited to solar
physics (Petrovay 1999). As in the turbulent cascade the diffusivity increases
for larger spatial scales, the transport of passive scalars by inertial-range
turbulence belongs to the subclass of {\it superdiffusive} processes.

The work presented here is a first exploration of the effect of superdiffusion
on the chemistry of molecular clouds. Our approach is described in Section~2. In
order to get a first impression of the importance of superdiffusive effects, we
neglect the intricacies of the chemical reaction network and treat their net
effect as a simple parametric source term in the diffusion equation. As
anomalous diffusivity cannot be described by any partial differential equation
in physical space, we are forced to work in Fourier space, with $D=D(k)$ a
function of the wavenumber $k$. To alleviate the ensuing difficulties, in
Section~3.1 we first consider a case where the density of the main component is
uniform throughout the cloud. A semianalytical solution is possible in this
case; the results indicate that there are significant and nontrivial differences
between the diffusive and superdiffusive cases.  Section~3.2 then considers what
changes are to be expected if the assumption of a uniform cloud is dropped.
Section~4 concludes the paper.

\section{Formulation of the problem}

Consider a spherically symmetric molecular cloud, with $r$ the radial
coordinate. Let $n_\mathrm{\scriptsize H_2}(r)$ denote the number density of
hydrogen, $n_i(r,t)$ the number density and $f_i={n_i \over n_{H_2}}$ the
fractional abundance of a tracer molecule $i$. On the basis of simple
mixing-length arguments (cf.\ XAL95), if turbulence is characterized by a
single dominant scale $l$ and velocity $v$, the diffusion equation for the
tracer reads
\begin{eqnarray} 
  \partial_t n_i(r,t) &=& D\nabla [\nabla  n_i(r,t) - f_i(r,t) 
  \nabla n_\mathrm{\scriptsize H_2}(r)] \\
  &&+ S_i(r,t) \nonumber , 
\end{eqnarray}
where $S_i(r,t)$ is the source/sink term, representing the chemistry reaction
scheme, and $D=\langle lv\rangle$ is the diffusion  coefficient. In previous
work the fiducial values $v\sim 1\,$km/s and $l\sim 0.1$ pc were used, yielding
diffusivities on the order of $10^{22}$--$10^{23}\,$cm$^2$/s.

As here we focus on the diffusive term, the chemical source term will be
represented very roughly as a simple relaxation term
\begin{equation} 
  S(r,t) = \frac{n_0(r) - n(r,t)}{ \tau_c }
\end{equation}
throughout this paper. Here $n_0(r)$ is the diffusionless equilibrium solution,
while $\tau_c$ is a relaxation timescale on which the solution would converge
to  $n_0(r)$, were there no diffusive effects. On the basis of the results
presented in XAL95 the value of $\tau_c$ can be estimated to lie in the range
$10^5$--$10^6$ years.

\begin{figure}{!t}
\includegraphics[width=80mm,height=50mm]{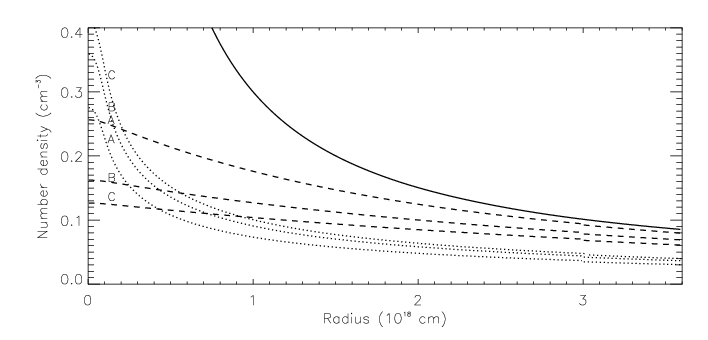}
\caption{Number density as a function of the radius in a uniform density cloud
for a tracer whose distribution scales as $n_0\sim 1/r$ in the nondiffusive case
(solid line).  Dashed line: normal diffusive case;  dotted line: superdiffusive
case. The parameters are:  $\mathrm{\tau_c=10^6 yrs}$, and D = $\mathrm{4\times
10^{22}}$ (A); $\mathrm{10^{23}}$ (B); $\mathrm{D=1.6\times 10^{23} cm^2s^{-1}}$
(C).}
\label{label1}
\end{figure}

\begin{figure}{!t}
\includegraphics[width=80mm,height=50mm]{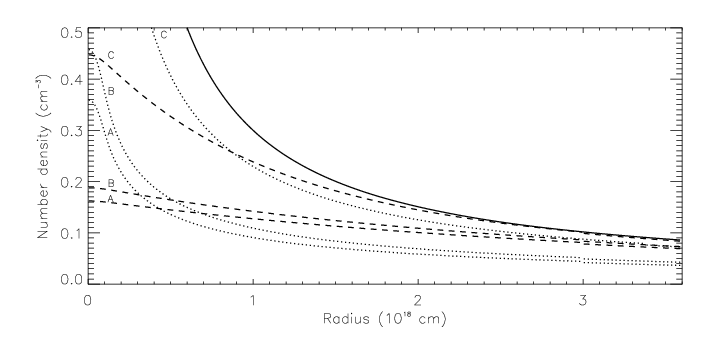}
\caption{Same as in Fig.~1. but the parameters are: $\mathrm{D=10^{23}
cm^2s^{-1}}$, and $\mathrm{\tau_c}$ = $\mathrm{10^6}$ (A);  $\mathrm{7.6\times
10^5}$ (B); $\mathrm{1.2\times 10^5 yrs}$ (C).}
\label{label1}
\end{figure}

\begin{figure}{!t}
\includegraphics[width=80mm,height=50mm]{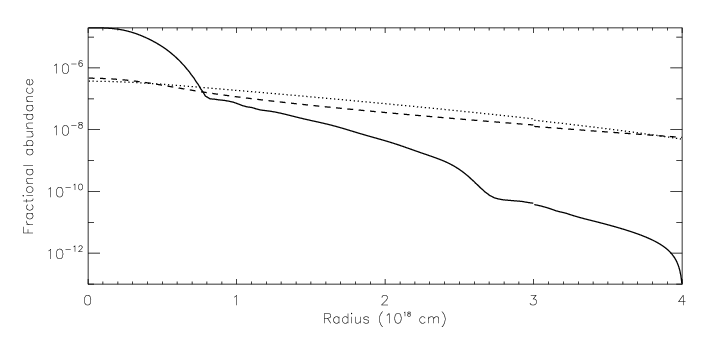}
\caption{Fractional abundance of ${\mathrm{O_2}}$ as a function of radius in a
uniform density cloud, with $D=10^{23}\,$cm$^2/$s, $\tau_c=10^6$ yrs. 
Solid line: nondiffusive case, adopted from XAL95.  
Dashed line: normal diffusive case;  dotted line: superdiffusive case.}
\label{label1}
\end{figure}

In a turbulent cascade regime where turbulence cannot be characterized by a
single dominant scale anymore, equation (1) does not hold. However, its Fourier
transform, with the appropriate scale-dependent diffusivity value $D(k)$ 
will still hold for each Fourier component of the passive scalar
field. For the wavenumber dependence of the diffusivity we use the Kolmogorov
form $D(k)=D_0(k/k_0)^{-4/3}$ throughout this paper, $D_0$ being the value of the
diffusivity at the wavenumber $k_0$, corresponding to the fiducial scale $l$ 
used in previous work.

\section{Results}

\subsection{Uniform density cloud}

The Fourier transform of equation (1) takes a particularly simple form if the
main constituent (i.e. H$_2$) is distributed uniformly. Equation (1) then
simplifies to 

\begin{equation} \partial_t n(r,t)  = D \nabla^2 n(r,t) + S(r,t) \end{equation}
The spatial Fourier transform of this is
\begin{equation} 
\partial_t \hat{n}(k,t) = - D(k) k^2 \hat{n}(k,t) + \hat{S}(k,t) 
\end{equation}
In the stationary case the solution is
\begin{equation} \hat{n}(k,t) = {\hat{n}_0(k,t) \over 1 + \tau_c D(k) k^2}.
\end{equation}
Transforming $\hat{n}(k,t)$ back to physical space we obtain the equilibrium
distribution of the tracer.

To carry out these transforms in the spherical geometry at hand, we recall the
theorem that an $n$ dimensional Fourier transform can be replaced a one
dimensional Hankel transform, if the transformable function depends only on $r
= \sqrt{\sum{}x_i^2}$ (Sneddon 1951). In three dimensions the transformation
formulae are:
\begin{equation} 
  \hat{n}(k) = \int_0^{\infty}{r^{3 \over 2}n(r)J_{1 \over 2}(kr)dr} 
\end{equation} 
\begin{equation} 
  n(r) = \int_0^{\infty}{k^{3 \over 2}\hat{n}(k)J_{1 \over 2}(kr)dk}.
\end{equation}
where $J_{1 \over 2}$ is the Bessel function of order $\frac 12$. These Hankel
transforms can only be computed analytically for some special cases, such as a
Gaussian profile for $n_0(r)$ (cf.\ Mar\-schal\-k\'o 2006). For the more
realistic cases presented below, equations (6) and (7) were evaluated
numerically.

\begin{figure}
\includegraphics[width=40mm,height=47mm]{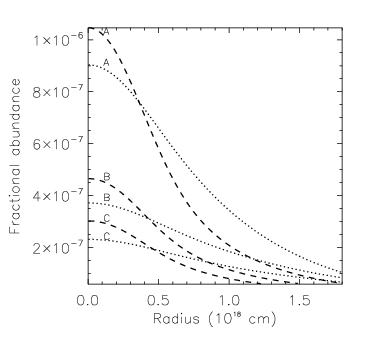}
\includegraphics[width=40mm,height=47mm]{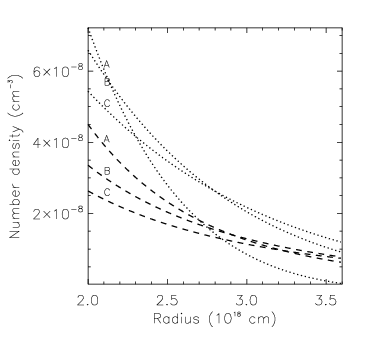}
\caption{Same as in Fig. 3. but the inner and outer parts of the cloud are shown
separately. The values of parameters are: $\mathrm{\tau_c=10^6 yrs}$,~D =
$\mathrm{4\times 10^{22}}$ (A), $\mathrm{10^{23}}$ (B) and $\mathrm{D=1.6\times
10^{23} cm^2s^{-1}}$ (C).}
\label{label1}
\end{figure}

In figures 1 and 2 we present analytical solutions assuming $n_0(r)\sim
r^{-1}$, a very rough analytical approximation for the actual profiles. It is
apparent that there is a significant difference between the diffusive and
superdiffusive case. In the superdiffusive case the concentration of the tracer
in the cloud core is much more pronounced, as a consequence of the reduced
diffusivity at small scales. Interestingly, an overall increase of the
diffusivity $D$ leads to an increase in the central number density of the
tracer in the diffusive case, while it leads to a decrease in the
superdiffusive case (though this trend reverses at large values of $r$).

Next, instead of the generic $1/r$ distribution we consider one particular
tracer, the oxygen molecule. The form of $n_0(r)$ for this molecule is taken
from XAL95. Figure 3 shows the fractional abundance of
O$_2$ as a function of the radius, still assuming a uniform H$_2$
distribution. The diffusive and superdiffusive cases are again significantly
different, but this distinction is better seen in the non-logarithmic plot
(Figure 4). In contrast to the $1/r$ case with its singular chemical source
function, in the superdiffusive solutions the tracer is now {\it less} 
concentrated to the core than with Fickian diffusion.

\begin{figure}
\includegraphics[width=80mm]{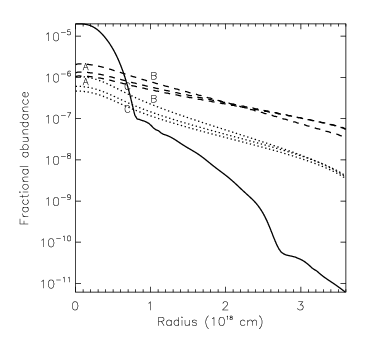}
\caption{Comparison of the O$_2$ fractional abundance profiles computed assuming
normal (Fickian) diffusion with uniform (dotted) and non-uniform (dashed) cloud
density.
Case A: $D=10^{23}\,$cm$^2$/s, $\tau_c=7.6\times 10^5$ yrs.
Case B: $D=4\times 10^{22}\,$cm$^2$/s, $\tau_c=10^6$ yrs.
Case C: $D=10^{23}\,$cm$^2$/s, $\tau_c=10^6$ yrs. 
Solid: initial profile $n_0(r)$.}
\label{label1}
\end{figure}

\begin{figure}
\includegraphics[width=80mm]{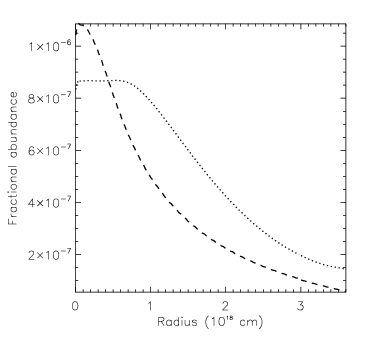}
\caption{Fractional abundance of O$_2$ as a function of radius in a cloud of
non-uniform density, with $D=10^{23}\,$cm$^2$/s, $\tau_c=7.6\times 10^5$ yrs.
Dashed line: normal diffusion. Dotted line: the expected abundance profile with
superdiffusion, obtained by the rescaling method described in the text.}
\label{label1}
\end{figure}

\subsection{Non-uniform cloud}

The non-uniformity of the cloud was in fact implicitly already taken into
account in the previous subsection, as $n_0(r)$ was taken from non-uniform cloud
models. To make our approach more consistent, it would be necessary to take into
account this non-uniformity also explicitly in equation (1). In this more
general case the problem becomes mathematically much more complicated. In order
to get a first impression of the importance of the inhomogeneity, here we first
compare the results obtained with and without explicitly treating non-uniformity
for the case of normal diffusion. For these calculations the density profile
$n_\mathrm{\scriptsize H_2}(r)$ was taken from XAL95, as above.

For the solution of equation (1) a time-relaxation method was used with a finite
difference scheme first order accurate in time and second order accurate in
space, on a grid of 512 points evenly distributed between
$r_\mathrm{\scriptsize{in}}=4\times 10^{15}$ and 
$r_\mathrm{\scriptsize{out}}=4\times 10^{18}\,$cm. At the inner boundary of our
domain the boundary condition was $\partial_r
n(r=r_\mathrm{\scriptsize{in}},t)=0$, while at the outer boundary the number
density was set to zero $n(r=r_\mathrm{\scriptsize{out}},t)=0$. Starting from
the same initial conditions as in the previous case, the solution was allowed to
evolve in time. Comparing the solutions obtained in the uniform case to the
exact stationary solution obtained from equation (5) we found that satisfactory
agreement is reached within one characteristic time $\tau_c$.

The results are plotted in figure~5. The effect of non-uniformity is clearly to
increase the tracer abundance, especially in the outer parts of the cloud.
Nevertheless, varying the parameters of the problem ($D_0$, $\tau_c$) the
resulting effects are qualitatively similar and quantitatively comparable to
those found in the uniform case. This encourages us to attempt to characterize
the effect of superdiffusion in the nonuniform case by rescaling the uniform
superdiffusive solution for the fractional abundance (dotted line in figure 3)
by the ratio of the nonuniform and uniform solutions in the normal diffusive
case. The resulting abundance curve is plotted in figure 6. Just as in the
uniform case (figure 3), superdiffusion reduces the O$_2$ abundance in the cloud
core and increases it in the envelope.

\section{Conclusion}

In this paper we considered to what extent and in what sense superdiffusion
modifies molecular abundances in interstellar clouds. The results clearly
demonstrate that the effect is quite significant, abundance values at a given
radius being modified by order of unity factors. The sense and character of this
influence is highly nontrivial.

In order to make a meaningful comparison with observational data possible, this
model needs to be developed further to include a realistic treatment of the
chemical reactions and to include a more rigorous calculation of the effect of
non-uniformity of the cloud in the diffusion equation. Work in this direction is
in progress.

\acknowledgements
This research was supported by the Theoretical Institute for
Advanced Research in Astrophysics (TIARA) operated under Academia Sinica and the
National Science Council Excellence Projects program in Taiwan administered
through grant number NSC95-2752-M-007-006-PAE, as well as by the Hungarian
Science Research Fund (OTKA) under grant no.\ K67746.

\end{document}